\documentclass[onecolumn,11pt,aps,nofootinbib,superscriptaddress,tightenlines]{revtex4}

\usepackage{graphics}
\usepackage{dsfont}
\usepackage{amsmath}
\usepackage{amssymb}
\usepackage{amsthm}
\usepackage{graphicx}
\usepackage{mathrsfs}
\usepackage{units}
\usepackage[usenames,dvipsnames]{color}

\usepackage[usenames,dvipsnames]{color}

\newcommand{\abs}[1]{\ensuremath{|#1|}}

\newcommand{\Abs}[1]{\ensuremath{\left|#1\right|}}

\newcommand{\norm}[2]{\ensuremath{|\!|#1|\!|_{#2}}}

\newcommand{\Norm}[2]{\ensuremath{\left|\!\left|#1\right|\!\right|_{#2}}}

\newcommand{\tr}{\textnormal{tr}}
\newcommand{\trace}[1]{\ensuremath{\tr (#1)}}
\newcommand{\Trace}[1]{\ensuremath{\tr \left( #1 \right)}}

\newcommand{\ptr}[1]{\textnormal{tr}_{#1}\,}
\newcommand{\ptrace}[2]{\ensuremath{\ptr{#1} (#2)}}

\newcommand{\idx}[2]{{#1}_{#2}}

\newcommand{\ket}[1]{| #1 \rangle}
\newcommand{\keti}[2]{| #1 \rangle_{#2}}

\newcommand{\proj}[2]{| #1 \rangle\!\langle #2 |}
\newcommand{\proji}[3]{| #1 \rangle\!\langle #2 |_{#3}}

\renewcommand{\d}[1]{\ensuremath{\textnormal{d}#1}}

\newcommand{\idi}[1]{\ensuremath{\idx{\mathds{1}}{#1}}}
\newcommand{\idA}{\idi{A}}

\newcommand{\opid}{\ensuremath{\mathcal{I}}}
\newcommand{\opidi}[1]{\ensuremath{\idx{\mathcal{I}}{#1}}}

\newcommand{\linops}[1]{\ensuremath{\mathcal{L}(#1)}}
\newcommand{\hermops}[1]{\ensuremath{\mathcal{L}^\dagger(#1)}}
\newcommand{\posops}[1]{\ensuremath{\mathcal{P}(#1)}}

\newcommand{\normstates}[1]{\ensuremath{\mathcal{S}_{=}(#1)}}
\newcommand{\subnormstates}[1]{\ensuremath{\mathcal{S}_{\leq}(#1)}}

\newcommand{\cD}{\mathcal{D}}
\newcommand{\cE}{\mathcal{E}}
\newcommand{\cF}{\mathcal{F}}
\newcommand{\cG}{\mathcal{G}}
\newcommand{\cH}{\mathcal{H}}

\newcommand{\cT}{\mathcal{T}}

\newcommand{\rhoB}{\ensuremath{\idx{\rho}{B}}}

\newcommand{\rhoAB}{\ensuremath{\idx{\rho}{AB}}}

\newcommand{\rhoAR}{\ensuremath{\idx{\rho}{AR}}}

\newcommand{\rhoR}{\ensuremath{\idx{\rho}{R}}}

\newcommand{\sigmaB}{\ensuremath{\idx{\sigma}{B}}}

\newcommand{\zetaR}{\ensuremath{\idx{\zeta}{R}}}

\newcommand{\cTAE}{\ensuremath{\idx{\cT}{A \ensuremath{\rightarrow} B}}}
\newcommand{\HA}{\ensuremath{\idx{\cH}{A}}}
\newcommand{\HB}{\ensuremath{\idx{\cH}{B}}}

\newcommand{\eps}{\varepsilon}

\newcommand{\chh}[5]{\ensuremath{H_{#1}^{#2}(#3|#4)_{#5}}}

\newcommand{\chmin}[3]{\chh{\textnormal{min}}{}{#1}{#2}{#3}}
\newcommand{\chmineps}[3]{\chh{\textnormal{min}}{\varepsilon}{#1}{#2}{#3}}
\newcommand{\chmineeps}[4]{\chh{\textnormal{min}}{#1}{#2}{#3}{#4}}

\newcommand{\EH}[1]{\ensuremath{\underset{\mathbb{U}}{\textnormal{\large{$\mathbb{E}$}}}\:#1}}
\newcommand{\ED}[1]{\ensuremath{\underset{\cD}{\textnormal{\large{$\mathbb{E}$}}}\:#1}}
\theoremstyle{plain}
\newtheorem{lemma}{Lemma}
\newtheorem{theorem}{Theorem}
\newtheorem{corollary}{Corollary}
\newtheorem{remark}{Remark}
\theoremstyle{definition}
\newtheorem{definition}{Definition}

\begin{document}
\title{Decoupling with unitary approximate two-designs}

\author{Oleg Szehr}
\email{o.dim.qit@googlemail.com}
\affiliation{Zentrum Mathematik, Technische Universit\"{a}t M\"{u}nchen, 85748 Garching, Germany}
\affiliation{Institute for Theoretical Physics, ETH Zurich, 8093 Zurich, Switzerland}

\author{Fr\'{e}d\'{e}ric Dupuis}
\affiliation{Department of Computer Science, Aarhus University, 8200 Aarhus N, Denmark}
\affiliation{Institute for Theoretical Physics, ETH Zurich, 8093 Zurich, Switzerland}

\author{Marco Tomamichel} 
\affiliation{Centre for Quantum Technologies, National University of Singapore, Singapore 117543}
\affiliation{Institute for Theoretical Physics, ETH Zurich, 8093 Zurich, Switzerland}

\author{Renato Renner}
\affiliation{Institute for Theoretical Physics, ETH Zurich, 8093 Zurich, Switzerland}

\date{\today}
\begin{abstract}
Consider a bipartite system, of which one subsystem, $A$, undergoes a physical evolution separated from the other subsystem, $R$. One may ask under which conditions this evolution destroys all initial correlations between the subsystems $A$ and $R$, i.e.~\emph{decouples} the subsystems.
A quantitative answer to this question is provided by \emph{decoupling theorems}, which have been developed recently in the area of quantum information theory. This paper builds on preceding work, which shows that decoupling is achieved if the evolution on $A$ consists of a typical unitary, chosen with respect to the Haar measure, followed by a process that adds sufficient decoherence. Here, we prove a generalized decoupling theorem for the case where the unitary is chosen from an approximate two-design. A main implication of this result is that decoupling is physical,
in the sense that it occurs already for short sequences of random two-body interactions, which can be modeled as efficient circuits. 
Our decoupling result is independent of the dimension of the $R$ system, which shows that approximate 2-designs are appropriate  for decoupling even if the dimension of this system is large.
\end{abstract}
\maketitle
\section{Introduction}
\label{intro}
Consider a joint quantum system, consisting of subsystems $A$ and $R$. We say that $A$ is \emph{decoupled} from $R$ if the joint state $\tau_{AR}$ has product form $\tau_A \otimes \tau_R$. Operationally, this means that the probability distributions obtained upon measuring the $A$ and $R$ systems are statistically independent. In this work, we are interested in processes acting locally on system $A$, which may initially be correlated to $R$, such that $A$ ends up being decoupled from $R$.

Processes that decouple a system $A$ from $R$ play an important role in various information theoretic applications. Examples abound in the area of quantum Shannon theory: state merging \cite{Merging} and state transfer~\cite{Mother}. Other important theorems, such as the best known achievable rates for sending quantum information through a quantum channel~\cite{Capacity}, can be proven concisely via decoupling. Moreover, arguments referring to decoupling have been used in a physical context and, for example, deepened our insight into the black hole information paradox~\cite{BLACK} and the role of negative conditional entropies in thermodynamics~\cite{ThermoNeg}. 

In~\cite{Fred:PHD}, a decoupling theorem has been derived that generalizes the previous decoupling theorems used in the aforementioned work. There one considers a situation where a subsystem $A$ of a joint system $AR$ undergoes an evolution while $R$ is left unchanged. The mapping describing the evolution of $A$ is conceptually split into two parts: a unitary followed by an arbitrary trace-preserving and completely positive map $\cT=\cTAE$. The Decoupling Theorem of \cite{Fred:PHD} (see also \cite{DTHM}) states that if an initial state $\rhoAR$ and a process $\cT$ are fixed and the unitary is taken either from the Haar measure or from a two-design~\cite{dankert09}, then the expected distance of the resulting state from a decoupled state is bounded in terms of entropic quantities:
$$\EH{\big\| \mathcal{T}\big((\idx{U}{A}\! \otimes\! \idi{R}) \rhoAR (\idx{U}{A}\! \otimes\! \idi{R})^\dagger\big) - \idx{\omega}{B} \otimes \idx{\rho}{R} \big\|}_1\leq
2^{-\frac{1}{2} H_{\min}(A'|B)_{\omega} - \frac{1}{2} H_{\min}(A|R)_{\rho} } .$$

Here the operator $\omega$ only depends on the map $\cTAE$ and, in particular, is independent of the chosen input state, $\rhoAR$. The min-entropy, $H_{\textnormal{min}}(A|R)_{\rho}$, (cf.~Definition~\ref{conditionalminentropy} below) quantifies the uncertainty an observer with access to $R$ has about the $A$ subsystem prior to the decoupling operation. The quantity $H_{\textnormal{min}}(A'|B)_{\omega}$ measures how well the mapping $\cTAE$ conserves correlations. It quantifies the uncertainty of an observer with access to the output subsystem $B$ about a copy $A'$ of the input state space, after the map $\cTAE$ is applied to a maximally entangled state on $AA'$. The min-entropy can be seen as a generalization of the well-known von Neumann entropy in the following sense. If a smoothed version of the min-entropy (cf.~Definition~\ref{def:smooth-min-entropy}) is evaluated for $n$ identical copies of the same state then in the asymptotic limit of large $n$ it reduces to the von Neumann entropy (cf.~Equation~\ref{QAEP}). Thus an important special case of the above relation arises when we consider the limit of a large number of identical copies of states, $\rhoAR$, and channels, $\cTAE$, applied to them. In this scenario the subsystems decouple if 
\begin{align*}
H(A'|B)_{\omega}+H(A|R)_{\rho}>0
\end{align*}
holds for the conditional von Neumann entropies of $\omega$ and $\rho$. Roughly, this inequality establishes a condition on the correlation in the initial state $\rho_{AR}$ and the \lq\lq{}decoupling power\rq\rq{} of the map $\cTAE$, which is sufficient for decoupling. Suppose for instance that $\rho_{AR}$ contains strong quantum correlations such that $H(A|R)_{\rho}$ is negative, then decoupling occurs if $\cTAE$ can destroy this correlation, that is $H(A'|B)_{\omega}$ is large enough for the above to hold. (See \cite{NielsenChuang} for a general introduction of negative conditional entropies and \cite{ThermoNeg} for their meaning in thermodynamics. A detailed discussion of sufficiency and necessity of the above condition for decoupling can be found in \cite{DTHM}.)

Often $\cTAE$ is chosen in a specific way. For example, in order to obtain the Fully Quantum Slepian-Wolf (FQSW) Theorem \cite{Mother}, it suffices to consider the case where $\cTAE$ is the partial trace. Another special case is state merging~\cite{Merging}, where $\cTAE$ represents a measurement of the $A$ system. In the FQSW scenario, the above inequality is known to be tight~\cite{Mother}.

In this paper we analyze whether decoupling occurs in a typical physical process. For this purpose, we generalize the decoupling theorem above to the case when the random unitary is taken from an \textit{approximate} two-design instead of a two-design.
Our discussion of approximate two-designs is motivated by the fact that, as opposed to exact two-designs such as the Clifford group \cite{datahiding,duer,gottesman-thesis}, approximate two-designs emerge in various realistic models of physical systems. In particular, approximate two-designs can be used to model a typical quantum mechanical evolution of an $A$ subsystem that is governed by two-particle interactions. More precisely, we follow the lines of~\cite{RQC} and model the internal dynamics of the $A$ subsystem in terms of a random quantum circuit and address the question of how well these dynamics decouple. We show that the quality of decoupling does not depend on the dimensions of the channel output $B$ and the reference system $R$ and prove that decoupling is physical, in the sense that it occurs already for short sequences of random two-body interactions even if $R$ is large\footnote{Note that it follows straight from continuity that approximate two-designs can be used for decoupling with an error depending on the approximation and the dimension of the physical system. However, in a physical scenario the dimensions of the channel output B and the reference system R can be large or unknown, which motivates the more elaborate analysis we provide in this article.}. Moreover, our decoupling results open the door to a more efficient implementation of operational tasks such as state transfer and state merging, since one might expect good approximate two-designs to outperform exact two-designs in terms of circuit complexity\footnote{Note that the circuit complexity of the exact two-design given by the Clifford group is quadratic, as shown in~\cite{gottesman-thesis}.}.

We note that the result achieved here has a (semi-) classical analogue, which is used, for instance in quantum cryptography for a task called \emph{privacy amplification}. Here the system $A$ is a classical random variable that is correlated with a quantum memory, $R$, held by an adversary. The goal is to extract randomness from $A$ which is private, i.e.~uncorrelated to the adversary's data $R$. This can be achieved by two-universal hash functions~\cite{carter79}, which replace the unitary two-design used in the Decoupling Theorem~\cite{rennerkoenig05}. An extension to \emph{almost} two-universal hash functions is already known in this classical scenario~\cite{TSSR10}. Our work can be seen as a fully quantum version of this result.

In this paper, we consider finite-dimensional systems only. However, the analogous task of privacy amplification described above has recently been extended to the case where the adversary holds an infinite-dimensional system~\cite{Furrer} or a general von Neumann algebra of observables~\cite{BFS}. The fact that our decoupling results do not involve the dimension of the system held by the adversary (and the dimension of the channel output) suggests that a similar generalization is also possible for decoupling.

The remainder of the paper is organized as follows. In Section~\ref{prel}, we introduce the mathematical framework used to derive our main technical results, which are presented in Section~\ref{decoup}. Finally, in Section~\ref{decoup:anal} we apply our results to analyze decoupling in a physical context.

\section{Preliminaries}
\label{prel}
\subsection{Notation}
\label{prel:not}
Let $\cH$ be a finite dimensional, complex Hilbert space. The set of linear operators on $\cH$ will be denoted by $\linops{\cH}$, the set of Hermitian operators by $\hermops{\cH}$ and the set of positive-semidefinite operators is given by $\posops{\cH}$. The set of quantum states is given by $\normstates{\cH} := \{\rho\in\posops{\cH} \mid \tr\, \rho = 1\}$ and the set of sub normalized quantum states is $\subnormstates{\cH} := \{\rho\in\posops{\cH} \mid \tr\,\rho \leq 1\}$. For the Lie group of unitary matrices we write $\mathbb{U}$. A subscript letter following some mathematical object denotes the physical system to which it belongs. However, when it is clear which systems are described we might drop the subscripts to shorten the notation.

Bipartite systems $AB$ are represented by a tensor product space $\idx{\cH}{A}\otimes\idx{\cH}{B}=:\idx{\cH}{AB}$. We will denote by $\idi{A}$ the identity operator on $\HA$ and by $\idx{\pi}{A}:= \idi{A} / \idx{d}{A}$ the completely mixed state on $A$, where $d_A = \dim \cH_A$. Linear maps from $\linops{\idx{\cH}{A}}$ to $\linops{\idx{\cH}{B}}$ will be denoted by calligraphic letters, e.g.~$\idx{\cT}{A\ensuremath{\rightarrow}B}$. Quantum operations are in one to one correspondence with the trace preserving  completely positive maps (TPCPMs). The TPCPM we will encounter most often is the partial trace (over the system $B$), denoted $\ptrace{B}{\cdot}$, which is defined to be the adjoint mapping of $\idx{\mathcal{T}}{A\rightarrow AB}(\idx{\xi}{A}) = \idx{\xi}{A} \otimes\idi{B}$ for $\idx{\xi}{A}\in\hermops{\HA}$ with respect to the Schmidt scalar product 
$\langle A,B\rangle \ := \ \tr(A^{\dagger}B)$. This means $\tr( (\idx{\xi}{A}\otimes\idi{B})\idx{\zeta}{AB}) = \tr(\idx{\xi}{A}\: \ptrace{B}{\idx{\zeta}{AB}})$ for any $\idx{\zeta}{AB}\in\hermops{\idx{\cH}{AB}}$.
Given a multipartite state $\idx{\xi}{AB}$, we write $\idx{\xi}{A} := \ptr{B}{\idx{\xi}{AB}}$ for the reduced density operator on $A$ and $\idx{\xi}{B} := \ptr{A} \xi_{AB}$, respectively, on $B$.

For isomorphic $\cH_A$ and $\cH_{A'}$, we denote by $\idx{\Phi}{AA'}$ the completely entangled state on $AA'$, i.e.~$\idx{\Phi}{AA'} := \proj{\Phi}{\Phi}_{AA'}$, where $\ket{\Phi}_{AA'}:= \sum_i \keti{i}{A} \otimes \keti{i}{A'} / \sqrt{\idx{d}{A}}$ and $\{ \keti{i}{A} \}$ and $\{ \keti{i}{A'} \}$ form 
orthonormal bases. 
%
The swap operator $\cF$ on the bipartite space $\cH_{AA\rq{}}$ is defined as
$\cF := \sum_{i,j} \proji{i}{j}{A} \otimes \proji{j}{i}{A\rq{}}.$
It is not difficult to verify \cite{swap1,swap2} that this operator satisfies
$\tr(MN)=\tr((M \otimes N)\mathcal{F})$ for any $M$, $N\in \linops{\cH_A}$. We refer to this observation as the \emph{swap trick}. The Choi-Ja\-mio\l{}\-kow\-ski representation~\cite{CHOI,JAM} of $\cTAE\in\textnormal{Hom}(\linops{\HA},\linops{\HB})$ is given by the operator ${\idx{\omega}{A'B}:=(\cTAE\otimes\idx{\opid}{A'})(\idx{\Phi}{AA'})}$. Here, $\idx{\opid}{A'}$ denotes the operator identity on $A'$, which we will only write explicitly if it is not clear from context.

For any operator in $\xi\in\linops{\cH}$ we denote by $\norm{\xi}{1}$, $\norm{\xi}{2}$ and $\norm{\xi}{\infty}$ the Schatten 1, 2 and $\infty$-norms of $\xi$, respectively. These norms are invariant under conjugation with unitaries and satisfy $\norm{\xi}{\infty}\leq\norm{\xi}{2}\leq\norm{\xi}{1}$. We will furthermore use that, for any $A,B,C\in\linops{\cH}$ and any Schatten norm $\norm{\cdot}{}$, it holds that $\norm{ABC}{}\leq\norm{A}{\infty}\norm{B}{}\norm{C}{\infty}$ (see, e.g.~\cite{Bhatia}).

The metric induced on $\linops{\cH}$ via the Schatten 1-norm is $D(\rho,\sigma):= \| \rho - \sigma \|_1$.
Another measure of distance on $\posops{\cH}$ is the fidelity, $F(\rho,\sigma) := \|\sqrt{\rho} \sqrt{\sigma} \|_1$. 
We also require a norm for linear maps $\cTAE$ from $\linops{\idx{\cH}{A}}$ to $\linops{\idx{\cH}{B}}$.
Given such a map, its diamond norm is defined to be \cite{diamond-norm}:
$$\norm{\cTAE}{\diamond}\ :=\ \sup_{\cH_R}{\max_{\rhoAR\in\linops{\idx{\cH}{AR}}}{\frac{\norm{\cTAE({\rhoAR})}{1}}{\norm{\rhoAR}{1}}}}.$$
Note that the diamond norm is the dual of the well-known norm of complete boundedness \cite{Paulsen}.
\subsection{Smooth Entropies}
\label{prel:smoo}
Entropies are used to quantify the uncertainty an observer has about a quantum state. Moreover, conditional entropies quantify the uncertainty of an observer about one subsystem of a bipartite state when he has access to another subsystem. 
The most commonly used quantity is the von Neumann entropy.
Given a state $\rhoAB \in \normstates{\cH_{AB}}$, we denote by $H(A|B)_{\rho} := H(\rhoAB) - H(\rhoB)$ the von Neumann entropy of $A$ conditioned on $B$.

While the von Neumann entropy is appropriate for analyzing processes involving a large number of copies of an identical system, the smooth min-entropy is the relevant quantity when a single system is considered~\cite{Renner:PHD}. Its definition is based on the following quantity.

\begin{definition}[Min-Entropy \cite{Renner:PHD}]\label{conditionalminentropy}
Let $\rhoAB\in\subnormstates{\idx{\cH}{AB}}$, then the min-entropy of $A$ conditioned on $B$ of $\rhoAB$ is defined as
$$\chmin{A}{B}{\rho}\::=\: \max_{\idx{\sigma}{B}\in\normstates{\HB}}\sup\{\lambda\in\mathbb{R}\mid\rhoAB\leq2^{-\lambda}\idA\otimes\sigmaB\}.$$
\end{definition}
More precisely the smooth conditional min-entropy is defined as the largest conditional min-entropy one can get within a distance of at most $\varepsilon$ from $\rho$. Here closeness is measured with respect to the \emph{purified distance}, $P(\rho,\sigma)$, which is defined to be
$$P(\rho,\sigma):= \sqrt{1-\bar{F}(\rho,\sigma)^2},$$
where $\bar{F}(\rho,\sigma)$ is the \emph{generalized fidelity};
$\bar{F}(\rho,\sigma):= F(\rho,\sigma)+\sqrt{(1-\tr\,\rho)(1-\tr\,\sigma)}$ for $\rho,\sigma\in\subnormstates{\cH}$. In~\cite{TCR09} it is shown that $P$ constitutes a metric on $\subnormstates{\cH}$ and the following inequalities are derived 
\begin{align}
\frac{1}{2}\left\|\rho-\sigma\right\|_1+\frac{1}{2}\abs{\tr\,\rho-\tr\,\sigma}\leq P(\rho,\sigma)\leq\sqrt{\left\|\rho-\sigma\right\|_1+\abs{\tr\,\rho-\tr\,\sigma}}\label{lem:genfuchs}.
\end{align}
We say that $\rho$ is $\eps$-close to $\tilde{\rho}$, denoted $\tilde{\rho}\approx_{\eps}\rho$, if $P(\rho,\tilde{\rho})\leq\varepsilon$.
\begin{definition}[Smooth Min-Entropy \cite{Renner:PHD,TCR09}]\label{def:smooth-min-entropy}
Let $\varepsilon\geq0$ and let $\rhoAB\in\subnormstates{\idx{\cH}{AB}}$ with $\sqrt{\tr{\rho}}>\varepsilon$, then the \textit{$\varepsilon$-smooth\ min-entropy} of $A$ conditioned on $B$ of $\rhoAB$ is defined as
$$\chmineps{A}{B}{\rho}\:\:=\:\max_{\tilde{\rho}}\chmin{A}{B}{\tilde{\rho}},$$
where we maximize over all $\tilde{\rho}\approx_{\eps}\rho$.
\end{definition}
The fully quantum asymptotic equipartition property (QAEP) states that in the limit of an infinite number of identical states the smooth min-entropy converges to the von Neumann entropy~\cite{FQAEP}: Let $\rhoAB\in\normstates{\cH_{AB}}$, then
\begin{align}
\lim_{\eps\rightarrow0}\lim_{n\rightarrow\infty}\frac{1}{n}\chmineps{A^n}{B^n}{\rho^{\otimes n}}=\chh{}{}{A}{B}{\rho} \label{QAEP}.
\end{align}
In that sense, the smooth conditional min-entropy can be seen as a one-shot generalization of the von Neumann entropy.
\subsection{Approximate Two-Designs and Quantum Circuits}
\label{prel:circ}
Heuristically, a unitary two-design is a finite subset $\mathcal{D}$ of $\mathbb{U}$ that has the property that averaging any polynomial of degree $2$ over $\cD$ gives the same result as integrating this polynomial over $\mathbb{U}$ with respect to the Haar measure, $\d{U}$. 
\begin{definition}[Unitary $\delta$-approximate two-design \cite{DankertThesis,dankert09,RQC}]\label{Def:GwGh} \sloppypar{Let $\mathcal{D}=\{(p_i, U_i)\}_{i=1,...,n}$ be a set of pairs, where the $U_i$ are unitary matrices on a Hilbert space $\cH$ and the $p_i\geq0$ with $\sum_i p_i=1$ are probabilities.
We define the maps}
$$\mathcal{G}_W(\rho):=\sum_i{p_i U_i^{\otimes 2}\rho (U_i^\dagger)^{\otimes 2}} \quad \textrm{and} \quad
\mathcal{G}_H(\rho):=\int_{\mathbb{U}}{{U^{\otimes 2}\rho (U^\dagger)^{\otimes 2}}}\d{U} $$
for $\rho\in\linops{\cH^{\otimes 2}}$.
The set $\mathcal{D}$ is called a unitary two-design if $\mathcal{G}_W=\mathcal{G}_H$. Furthermore, $\mathcal{D}$ is called a $\delta$-approximate unitary two-design if $\norm{\mathcal{G}_W-\mathcal{G}_H}{\diamond}\leq\delta$.
\end{definition}
We will denote an integral over the unitary group with respect to the normalized Haar measure by ${\textnormal{\large{$\mathbb{E}$}}}_{\mathbb{U}}(\cdot)$ and an average over a unitary approximate two-design by ${\textnormal{\large{$\mathbb{E}$}}}_{\cD}(\cdot)$ for notational convenience.

For the applications that we are interested in, the most relevant approximate designs are generated by random quantum circuits \cite{RQC}. A quantum circuit is a set of wires on which gates are applied. Each wire corresponds to a qubit evolving in time, and each gate on the wire corresponds to some unitary operation being applied to the qubit. A $k$-qubit gate is given by an element of $\mathbb{U}(2^k)$. For us it will be sufficient to think of the circuit as a sequence of unitaries that are applied in a certain order: $W=W_t\cdot...\cdot W_2\cdot W_1$, where we call $t$ the time of the circuit. We call a set of gates \textit{universal} for $n$ qubits if any operation that can be performed on $n$ qubits can be approximated to arbitrary precision using operations from the universal gate set only.
\section{Decoupling with $\delta$-approximate Unitary Two-Designs}
\label{decoup}
We prove a decoupling theorem which applies to the general case where the evolution is described by a unitary chosen from a $\delta$-approximate two-design followed by an arbitrary physical process.

\begin{theorem}(Decoupling with $\delta$-approximate unitary two-designs)\label{dcpwthaldsg}
Let $\rhoAR\in\subnormstates{\idx{\cH}{AR}}$ be a subnormalized density operator and let $\cTAE$ be a linear map with Choi-Jamio\l{}kowski representation $\idx{\omega}{A'B}\in\subnormstates{\idx{\cH}{BA'}}$, then
\begin{align}
&\ED{\norm{\mathcal{T}((\idx{U}{A} \otimes \idi{R}) \:\rhoAR \: (\idx{U}{A}^\dagger \otimes \idi{R})) - \idx{\omega}{B} \otimes \idx{\rho}{R}}{1}}\nonumber\\
&\qquad\qquad\qquad\leq\sqrt{1+4\delta\idx{d}{A}^4}\ 2^{-\frac{1}{2}\:(\chmin{A'}{B}{\omega}\ +\ \chmin{A}{R}{\rho})},\nonumber
\end{align}
where $\cD$ constitutes a $\delta$-approximate two-design.
\end{theorem}

\begin{remark}
	It should be noted that the factor $d_A^4$ in the above formula can be compensated for by making $\delta$ accordingly small. See Section~\ref{decoup:anal} for a specific example, where the approximate two-design is created by a random circuit.
\end{remark}
\begin{remark}
 {Since the above decoupling formula does not involve the dimension factors $d_B$ and $d_R$ a $\delta$-approximate 2-design (with fixed $\delta$) yields decoupling even if one of these factors is intractably large.} 
\end{remark}

 {Note that Theorem~\ref{dcpwthaldsg} does not follow straight from a simple argument based on continuity. If exact 2-designs work in the sense of decoupling one expects that $\delta$-approximate 2-designs should work approximately. The error due to approximation depends on $\delta$ and, due to norm equivalence (compare also Lemma 2.2.14 in \cite{Low:PHD}), the dimension of the expression in the norm above. However, the upper bound of Theorem~\ref{dcpwthaldsg} does not involve the dimensions of the systems $B$ and $R$. Hence, it allows for the conclusion that decoupling can occur in a physical scenario, where the evolution of the $A$ subsystem is modeled as a (short) quantum circuit and the reference system $R$ potentially is large (see Section~\ref{decoup:anal}).}
We also remark that in the particular case of a perfect two-design, the proof of Theorem~\ref{dcpwthaldsg} includes a shorter derivation of the decoupling theorem for perfect two-designs as opposed to the original proof in~\cite{Fred:PHD,DTHM} (see Section~\ref{decoup:thm}).

The rest of this section is structured in four subsections. First, we prove a lemma that quantifies decoupling in terms of Schatten 2-norms. Then, in Section~\ref{decoup:thm}, we derive the decoupling formula for perfect two-designs using that lemma (see Theorem~\ref{Decouplthrm}). Section~\ref{decoup:2-dsg} is devoted to the derivation and analysis of the decoupling formula for general $\delta$-approximate two-designs (see Theorem~\ref{dcpwthaldsg}). And lastly, in Section~\ref{decoup:smoo} we reformulate the upper bound given by the decoupling formula for $\delta$-approximate two-designs in terms of smooth conditional min-entropies (see Theorem~\ref{wialtwo}). This enables us to make statements about independent, identically distributed states via the QAEP, Equation~\eqref{QAEP}.
\subsection{Decoupling with Schatten 2-Norms}
\label{decoup:lemma}
For a map $\cT\in$ $\textnormal{Hom}(\linops{\HA},\linops{\HB})$ with Choi-Jamio\l{}kowski representation $\idx{\omega}{A'B}\in\hermops{\idx{\cH}{BA'}}$ and an operator $\rhoAR\in\hermops{\idx{\cH}{AR}}$, we prove that
\begin{align}&\EH{ {\left\| \mathcal{T}((\idx{U}{A} \otimes \idi{R}) \: \rhoAR \: (\idx{U}{A}^\dagger \otimes \idi{R})) - \idx{\omega}{B} \otimes \idx{\rho}{R} \right\|}_2^2 }\nonumber\\
&=\frac{\idx{d}{A}^2}{\idx{d}{A}^2-1}\:{\left\|\idx{\rho}{AR}-\idx{\pi}{A}\otimes\rhoR\right\|}_2^2\: {\left\|\idx{\omega}{A'B}-\idx{\pi}{A'}\otimes\idx{\omega}{B}\right\|}_2^2.\label{decouplemmab}
\end{align}
For our application and the proof of \eqref{decouplemmab} it is convenient to reformulate the argument of the expectation value in a more symmetric way. We introduce the map $\idx{\cE}{\ensuremath{\tilde{A}\rightarrow}R}$, which we define to be the unique Choi-Jamio\l{}kowski preimage of the state $\rhoAR$ i.e.~$\idx{\cE}{\ensuremath{\tilde{A}\rightarrow}R}(\idx{\Phi}{A\ensuremath{\tilde{A}}})=\rhoAR$, where $\tilde{A}$ is just a copy of $A$. Note that $\cE$ is not trace-preserving in general. We can write for any unitary $\idx{U}{A}$:
\begin{align}
&\mathcal{T}((\idx{U}{A} \otimes \idi{R}) \ \rhoAR \  (\idx{U}{A}^\dagger \otimes \idi{R})) - \idx{\omega}{B} \otimes \idx{\rho}{R} \label{advxi}\nonumber\\
&=(\mathcal{T}\otimes\cE)((\idx{U}{A} \otimes \idi{\ensuremath{\tilde{A}}})\: \idx{\Phi}{A\ensuremath{\tilde{A}}}\:(\idx{U}{A}^\dagger \otimes \idi{\ensuremath{\tilde{A}}})) - (\cT\otimes\cE)(\idx{\pi}{A}\otimes\idx{\pi}{\ensuremath{\tilde{A}}})\\
&=(\cT\otimes\cE)((\idx{U}{A} \otimes \idi{\ensuremath{\tilde{A}}})\: \idx{\xi}{A\ensuremath{\tilde{A}}}\:(\idx{U}{A}^\dagger \otimes \idi{\ensuremath{\tilde{A}}}))\label{intxinew},
\end{align}
where we have introduced the \emph{decoupling operator} $\xi_{A\tilde{A}} := \Phi_{A\tilde{A}} - \pi_A \otimes \pi_{\tilde{A}}$. Equation \eqref{advxi}, uses the fact that an arbitrary map acting exclusively on the $A$ subsystem of $\idx{\Phi}{A\ensuremath{\tilde{A}}}$ commutes with any map that only acts on $\tilde{A}$. In Equation \eqref{intxinew} the linearity of the maps is used.
Analogously one has that
\begin{align*}
	\cE(\idx{\xi}{A\ensuremath{\tilde{A}}}) &= \rhoAR-\idx{\pi}{A}\otimes\rhoR & \cT(\idx{\xi}{A\ensuremath{\tilde{A}}}) &= \idx{\omega}{\ensuremath{\tilde{A}}B}-\idx{\pi}{\ensuremath{\tilde{A}}}\otimes\idx{\omega}{B}.
\end{align*}
Thus the stated result, Equation~\ref{decouplemmab}, can be rewritten equivalently in terms of the decoupling operator.
\begin{lemma}\label{decouplemma2}
Let $\idx{\xi}{A\ensuremath{\tilde{A}}}=\idx{\Phi}{A\ensuremath{\tilde{A}}}-\idx{\pi}{A}\otimes\idx{\pi}{\ensuremath{\tilde{A}}}$  and let $\cTAE\in\textnormal{Hom}(\linops{\HA},\linops{\HB})$ and $\idx{\cE}{\ensuremath{\tilde{A}\rightarrow}R}\in\textnormal{Hom}(\linops{\idx{\cH}{\ensuremath{\tilde{A}}}},\linops{\idx{\cH}{R}})$ be linear maps that preserve hermiticity, then
\begin{align}\EH{ \Norm{(\cT\otimes\cE)((\idx{U}{A} \otimes \idi{\ensuremath{\tilde{A}}})\:  \idx{\xi}{A\ensuremath{\tilde{A}}}\:  (\idx{U}{A}^\dagger \otimes \idi{\ensuremath{\tilde{A}}}))}{2}^2 }
=\frac{\idx{d}{A}^2}{\idx{d}{A}^2-1}\:{\left\|\cE(\idx{\xi}{A\ensuremath{\tilde{A}}})\right\|}_2^2\: {\left\|\cT(\idx{\xi}{A\ensuremath{\tilde{A}}})\right\|}_2^2.\nonumber
\end{align}
\end{lemma}

\begin{proof} We have that
\begin{align}
&\EH{\Norm{(\cT\otimes\cE)((\idx{U}{A} \otimes \idi{\ensuremath{\tilde{A}}})\:  \idx{\xi}{A\ensuremath{\tilde{A}}}\:   (\idx{U}{A}^\dagger \otimes \idi{\ensuremath{\tilde{A}}}))}{2}^2}\nonumber\\
&=\EH{{\Trace{(\cT\otimes\cE)((\idx{U}{A} \otimes \idi{\ensuremath{\tilde{A}}})\: \idx{\xi}{A\ensuremath{\tilde{A}}}\:  (\idx{U}{A}^\dagger \otimes \idi{\ensuremath{\tilde{A}}}))^2}}}\nonumber\\
&=\EH{{\Trace{(\cT\otimes\cE)^{\otimes2}\left((\idx{U}{A} \otimes \idi{\ensuremath{\tilde{A}}})^{\otimes2}\  (\idx{\xi}{A\ensuremath{\tilde{A}}})^{\otimes2}\   (\idx{U}{A}^\dagger \otimes \idi{\ensuremath{\tilde{A}}})^{\otimes2}\right)\:\idx{\cF}{BR}}}}\label{thswp}\\
&=\EH{{\Trace{\left((\idx{U}{A} \otimes \idi{\ensuremath{\tilde{A}}})^{\otimes2}\  (\idx{\xi}{A\ensuremath{\tilde{A}}})^{\otimes2}\   (\idx{U}{A}^\dagger \otimes \idi{\ensuremath{\tilde{A}}})^{\otimes2}\right)\:(\cT^\dagger)^{\otimes2}[\idx{\cF}{B}]\otimes(\cE^\dagger)^{\otimes2}[\idx{\cF}{R}]}}}.\label{dfndgr}
\end{align}
We introduced two further copies $A'$ and $\tilde{A}'$ of $A$ when using the swap trick in Equation \eqref{thswp}, i.e. $(\idx{\xi}{A\ensuremath{\tilde{A}}})^{\otimes2}\:=\:\idx{\xi}{A\ensuremath{\tilde{A}}}\otimes\idx{\xi}{A'\ensuremath{\tilde{A'}}}$.
In Equation \eqref{dfndgr} we used the definition of the adjoint of the mapping $(\mathcal{T}\otimes\cE)^{\otimes2}$ with respect to the Schmidt scalar product. We have from \cite{Fred:PHD}, Lemma 3.4, that
\begin{align*}
\EH{\left((\idx{U}{A})^{\dagger\otimes2}(\mathcal{T}^{\dagger})^{\otimes2}(\idx{\mathcal{F}}{B})(\idx{U}{A})^{\otimes2}\right)}
= \alpha \idi{AA'} + \beta\idx{\cF}{A},
\end{align*}
with the coefficients $\alpha$ and $\beta$ satisfying
\begin{align*}
\alpha =\trace{\idx{\omega}{B}^2}\left(\frac{\idx{d}{A}^2-\frac{\idx{d}{A}\:\Trace{\idx{\omega}{A'B}^2}}{\Trace{\idx{\omega}{B}^2}}}{\idx{d}{A}^2-1}\right)\quad\textnormal{and}\quad
\beta =\trace{\idx{\omega}{A'B}^2}\left(\frac{\idx{d}{A}^2-\frac{\idx{d}{A}\:\Trace{\idx{\omega}{B}^2}}{\Trace{\idx{\omega}{A'B}^2}}}{\idx{d}{A}^2-1}\right).
\end{align*}
Similar integrals were evaluated in the context of decoupling already in \cite{Merging}. Using the above we get
\begin{align}
&\EH{\Norm{(\cT\otimes\cE)((\idx{U}{A} \otimes \idi{\ensuremath{\tilde{A}}})\: \idx{\xi}{A\ensuremath{\tilde{A}}}\:  (\idx{U}{A}^\dagger \otimes \idi{\ensuremath{\tilde{A}}}))}{2}^2}\nonumber\\
&=\Trace{(\idx{\xi}{A\ensuremath{\tilde{A}}})^{\otimes2}\{ \alpha \idi{AA'} + \beta\idx{\cF}{A}\}\otimes(\cE^\dagger)^{\otimes2}[\idx{\cF}{R}]}\nonumber\\
&=\beta\:\Trace{(\idx{\xi}{A\ensuremath{\tilde{A}}})^{\otimes2}\ \idx{\cF}{A}\otimes(\cE^\dagger)^{\otimes2}[\idx{\cF}{R}]}\label{scndtlla}\\
&=\beta\:\Norm{\cE(\idx{\xi}{A\ensuremath{\tilde{A}}})}{2}^2\label{withbet}.
\end{align}
In Equation \eqref{scndtlla} we used that tracing out one of the subsystems $A$, $\tilde{A}$ of $\idx{\xi}{A\ensuremath{\tilde{A}}}$ gives the zero state. The last line above makes use of the definition of the adjoint of $\cE$, the swap trick and the definition of the Schatten 2-norm.
Rewriting $\beta$ we find that
\begin{align}
\beta &=\trace{\idx{\omega}{A'B}^2}\left(\frac{\idx{d}{A}^2-\frac{\idx{d}{A}\:\Trace{\idx{\omega}{B}^2}}{\Trace{\idx{\omega}{A'B}^2}}}{\idx{d}{A}^2-1}\right)\nonumber\\
&=\frac{\idx{d}{A}^2}{\idx{d}{A}^2-1}\:\Norm{\cT(\idx{\xi}{A\ensuremath{\tilde{A}}})}{2}^2\label{shbet2}.
\end{align}
Substituting this into Equation \eqref{withbet} yields
\begin{align*}
&\EH{\Norm{(\cT\otimes\cE)((\idx{U}{A} \otimes \idi{\ensuremath{\tilde{A}}})\:  \idx{\xi}{A\ensuremath{\tilde{A}}}\:   (\idx{U}{A}^\dagger \otimes \idi{\ensuremath{\tilde{A}}}))}{2}^2}\nonumber\\
&=\frac{\idx{d}{A}^2}{\idx{d}{A}^2-1}\:\Norm{\cT(\idx{\xi}{A\ensuremath{\tilde{A}}})}{2}^2\:\Norm{\cE(\idx{\xi}{A\ensuremath{\tilde{A}}})}{2}^2,
\end{align*}
which proves the lemma.
\end{proof}
\subsection{Decoupling with Perfect Two-Designs}
\label{decoup:thm}
In this subsection we show two additional lemmas that we require for the derivation of our main result, Theorem~\ref{dcpwthaldsg}. Taking these lemmas together with Lemma~\ref{decouplemma2}, we also obtain a concise derivation of the decoupling theorem for the Haar measure (cf. Theorem~\ref{Decouplthrm}).
\begin{lemma}\label{hoeldr}
Let $\idx{\xi}{BR}\in\hermops{\idx{\cH}{BR}}$ and let $\idx{\lambda}{BR}\in\normstates{\idx{\cH}{BR}}$ be invertible. Then
$$\norm{\idx{\xi}{BR}}{1}\leq\norm{\idx{\lambda}{BR}^{-\frac{1}{4}}\:\idx{\xi}{BR}\:\idx{\lambda}{BR}^{-\frac{1}{4}}}{2}.$$
\end{lemma}
\begin{proof}
The Lemma follows from an application of the H\"older-type inequality $\Norm{ABC}{1}\leq\Norm{\abs{A}^4}{1}^{\frac{1}{4}}\Norm{\abs{B}^2}{1}^{\frac{1}{2}}\Norm{\abs{C}^4}{1}^{\frac{1}{4}}$ (see, for example, \cite{Bhatia}), with $A=C=(\idx{\lambda}{BR})^{\frac{1}{4}}$ and $B=\idx{\lambda}{BR}^{-\frac{1}{4}}\:\idx{\xi}{BR}\:\idx{\lambda}{BR}^{-\frac{1}{4}}$.
\end{proof}
\begin{lemma}\label{minentr}
For any $\idx{\xi}{AR}\in\subnormstates{\idx{\cH}{AR}}$ there is $\idx{\zeta}{R}\in\normstates{\idx{\cH}{R}}$ with
$$\frac{1}{\tr{[\idx{\xi}{AR}]}}\Trace{((\idi{A}\otimes\zetaR^{-1/2})\idx{\xi}{AR})^2}\leq2^{-\chmin{A}{R}{\xi}}.$$
\end{lemma}
\begin{proof} Choose $\zetaR$ such that it saturates the bound in the definition of the $\idx{H}{\textnormal{min}}$-entropy. Without loss of generality $\zetaR$ is invertible (otherwise, redefine $R$ such that it corresponds to the support of $\rhoAR$). Then
\begin{align*}
\xi_{AR}&\leq2^{-\chmin{A}{R}{\xi}}\idi{A}\otimes\zetaR,
\end{align*}
which implies that there is $\zetaR$ with
\begin{align}
\sqrt{\xi_{AR}}\:(\idi{A}\otimes\zetaR^{-\frac{1}{2}})\xi_{AR}(\idi{A}\otimes\zetaR^{-\frac{1}{2}})\:\sqrt{\xi_{AR}}\leq2^{-\chmin{A}{R}{\xi}}\xi_{AR}\label{conprstr}.
\end{align}
Taking the trace on both sides of \eqref{conprstr} proves Lemma~\ref{minentr}.
\end{proof}

Before proving our main theorem, it will be useful for the sake of completeness to first state and prove the decoupling theorem of \cite{Fred:PHD} in the formulation which is given in~\cite{DTHM}:
\begin{theorem} (Decoupling Theorem, \cite{Fred:PHD})\label{Decouplthrm}
Let $\rhoAR\in\subnormstates{\idx{\cH}{AR}}$ be a subnormalized density operator and let $\cTAE$ be a linear map with Choi-Jamio\l{}kowski representation $\idx{\omega}{A'B}\in\subnormstates{\idx{\cH}{BA'}}$, then
$$\EH{ \norm{\mathcal{T}((\idx{U}{A} \otimes \idi{R}) \:\rhoAR \:(\idx{U}{A}^\dagger \otimes \idi{R})) - \idx{\omega}{B} \otimes \idx{\rho}{R}}{1} } \leq
2^{-\frac{1}{2} H_{\textnormal{min}}(A'|B)_{\omega} - \frac{1}{2} H_{\textnormal{min}}(A|R)_{\rho} }.$$
\end{theorem}
\begin{proof}
Note first that for a proof of Theorem~\ref{Decouplthrm} it suffices to show that
\begin{align}
\EH{ \norm{\mathcal{T}((\idx{U}{A} \otimes \idi{R}) \ \rhoAR \  (\idx{U}{A}^\dagger \otimes \idi{R})) - \idx{\omega}{B} \otimes \idx{\rho}{R}}{1}^2 } \leq
2^{-H_{\textnormal{min}}(A'|B)_{\omega} - H_{\textnormal{min}}(A|R)_{\rho} }\label{drs18}
\end{align}
holds and to apply the Jensen Inequality. To prove Equation \eqref{drs18}, we work with the integrand in terms of the decoupling operator (Lemma~\ref{decouplemma2}). We use Lemma~\ref{hoeldr} to bound the Schatten 1-norm of the integrand with the Schatten 2-norm. Introducing the positive and normalized operators $\idx{\sigma}{B}$ and $\idx{\zeta}{R}$, we have
\begin{align*}
&\Norm{(\cT\otimes\cE)((\idx{U}{A} \otimes \idi{\ensuremath{\tilde{A}}})\: \idx{\xi}{A\ensuremath{\tilde{A}}}\:(\idx{U}{A}^\dagger \otimes \idi{\ensuremath{\tilde{A}}}))}{1}\nonumber\\
&\leq\Norm{(\sigmaB\otimes\zetaR)^{-\frac{1}{4}}\left((\cT\otimes\cE)((\idx{U}{A} \otimes \idi{\ensuremath{\tilde{A}}})\: \idx{\xi}{A\ensuremath{\tilde{A}}}\:(\idx{U}{A}^\dagger \otimes \idi{\ensuremath{\tilde{A}}}))\right)(\sigmaB\otimes\zetaR)^{-\frac{1}{4}}}{2}.
\end{align*}
One can abbreviate the notation using the completely positive maps $\idx{\tilde{\mathcal{T}}}{A \ensuremath{\rightarrow} B}$ and $\idx{\tilde{\cE}}{\ensuremath{\tilde{A} \rightarrow} R}$ defining
\begin{align}
\tilde{\mathcal{T}}(\idx{\tau}{A\ensuremath{\tilde{A}}})\ &:=\ (\sigmaB \otimes \idi{\ensuremath{\tilde{A}}})^{-1/4} \mathcal{T}(\idx{\tau}{A\ensuremath{\tilde{A}}}) (\sigmaB \otimes \idi{\ensuremath{\tilde{A}}})^{-1/4}\qquad\ \forall\ \idx{\tau}{A\ensuremath{\tilde{A}}}\in\linops{\idx{\cH}{A\ensuremath{\tilde{A}}}},\label{intrtilt}\\
\tilde{\mathcal{E}}(\idx{\tau}{A\ensuremath{\tilde{A}}})\ &:=\ (\idi{A} \otimes \zetaR)^{-1/4} \cE(\idx{\tau}{A\ensuremath{\tilde{A}}})(\idi{A} \otimes \zetaR)^{-1/4}\qquad\ \ \forall\ \idx{\tau}{A\ensuremath{\tilde{A}}}\in\linops{\idx{\cH}{A\ensuremath{\tilde{A}}}}\label{intrtile},
\end{align}
and $\idx{\tilde{\omega}}{A'B}:=\tilde{\mathcal{T}}(\Phi_{AA\rq{}}),\  \tilde{\rho}_{AR}:=\tilde{\mathcal{E}}(\Phi_{A\tilde{A}})$, which yields
\begin{align}
&\EH{{\left\| (\cT\otimes\cE)((\idx{U}{A} \otimes \idi{\ensuremath{\tilde{A}}})\:\idx{\xi}{A\ensuremath{\tilde{A}}}\:(\idx{U}{A}^\dagger \otimes \idi{\ensuremath{\tilde{A}}}))\right\|}_1^2}\nonumber\\
&\leq\EH{{\left\| (\tilde{\cT}\otimes\tilde{\cE})((\idx{U}{A} \otimes \idi{\ensuremath{\tilde{A}}}) \:\idx{\xi}{A\ensuremath{\tilde{A}}}\:(\idx{U}{A}^\dagger \otimes \idi{\ensuremath{\tilde{A}}}))\right\|}_2^2}\label{traceterm}\\
&=\frac{\idx{d}{A}^2}{\idx{d}{A}^2-1}\:\Norm{\tilde{\cT}(\idx{\xi}{A\ensuremath{\tilde{A}}})}{2}^2\:\Norm{\tilde{\cE}(\idx{\xi}{A\ensuremath{\tilde{A}}})}{2}^2\nonumber.
\end{align}
By Equation \eqref{shbet2} we have that
\begin{align}
&\frac{\idx{d}{A}^2}{\idx{d}{A}^2-1}\:\Norm{\tilde{\cT}(\idx{\xi}{A\ensuremath{\tilde{A}}})}{2}^2\:\Norm{\tilde{\cE}(\idx{\xi}{A\ensuremath{\tilde{A}}})}{2}^2\nonumber\\
&=(1-\frac{1}{\idx{d}{A}^2})\:\trace{\idx{\tilde{\omega}}{A'B}^2}\:\trace{\idx{\tilde{\rho}}{AR}^2}\left(\frac{\idx{d}{A}^2-\frac{\idx{d}{A}\:\Trace{\idx{\tilde{\omega}}{B}^2}}{\Trace{\idx{\tilde{\omega}}{A'B}^2}}}{\idx{d}{A}^2-1}\right)\left(\frac{\idx{d}{A}^2-\frac{\idx{d}{A}\:\Trace{\idx{\tilde{\rho}}{R}^2}}{\Trace{\idx{\tilde{\rho}}{AR}^2}}}{\idx{d}{A}^2-1}\right)\nonumber\\
&\leq\frac{1}{\tr[\idx{\omega}{A'B}]}\trace{\idx{\tilde{\omega}}{A'B}^2}\:\frac{1}{\tr[\idx{\rho}{AR}]}\trace{\idx{\tilde{\rho}}{AR}^2}\label{uslm}.
\end{align}
In Equation \eqref{uslm} we used the Cauchy-Schwarz inequality (Lemma 3.5 in \cite{Fred:PHD}) to infer that both bracket terms are smaller than one. The derivation is valid for any positive and normalized operators $\idx{\sigma}{B}$ and $\zetaR$, therefore one can choose $\idx{\hat{\sigma}}{B}$ and $\idx{\hat{\zeta}}{R}$
such that they minimize the expression in \eqref{uslm}. An application of Lemma~\ref{minentr} then shows that
\begin{align*}
&\EH{ {\left\| \mathcal{T}((\idx{U}{A} \otimes \idi{R}) \:\rhoAR \:(\idx{U}{A}^\dagger \otimes \idi{R})) - \idx{\omega}{B} \otimes \idx{\rho}{R} \right\|}_1^2 }
\leq2^{-\chmin{A'}{B}{\omega}\ -\ \chmin{A}{R}{\rho}}.
\end{align*}
\end{proof}
\subsection{Decoupling with $\delta$-approximate Two-Designs}
\label{decoup:2-dsg}
This subsection is devoted to a proof of the core theorem of this paper:
\begin{proof}[Proof of Theorem~\ref{dcpwthaldsg}]
Due to the Jensen Inequality it suffices to show that
\begin{align}
&\ED{\norm{\mathcal{T}((\idx{U}{A} \otimes \idi{R}) \ \rhoAR \  (\idx{U}{A}^\dagger \otimes \idi{R})) - \idx{\omega}{B} \otimes \idx{\rho}{R}}{1}^2}\nonumber\\
&\leq\left(1+4\delta\idx{d}{A}^4\right)\ 2^{-\chmin{A'}{B}{\omega}-\chmin{A}{R}{\rho}}\label{mchop}
\end{align}
holds. To prove \eqref{mchop}, we proceed in a similar fashion to our proof of Theorem~\ref{Decouplthrm}.
As before, we introduce the map $\idx{\cE}{\ensuremath{\tilde{A}\rightarrow}B}$ which we define to be the unique Choi-Jamio\l{}kowski preimage of $\idx{\rho}{AR}$ and the state $\idx{\xi}{A\ensuremath{\tilde{A}}}=\idx{\Phi}{A\ensuremath{\tilde{A}}}-\idx{\pi}{A}\otimes\idx{\pi}{\ensuremath{\tilde{A}}}$ and write for any unitary:
\begin{align*}
\mathcal{T}((\idx{U}{A} \otimes \idi{R}) \rhoAR(\idx{U}{A}^\dagger \otimes \idi{R})) - \idx{\omega}{B} \otimes \idx{\rho}{R}
=(\cT\otimes\cE)((\idx{U}{A} \otimes \idi{\ensuremath{\tilde{A}}}) \idx{\xi}{A\ensuremath{\tilde{A}}}(\idx{U}{A}^\dagger \otimes \idi{\ensuremath{\tilde{A}}})).
\end{align*}
To upper bound the left-hand side of \eqref{mchop}, we apply Lemma~\ref{hoeldr}. We introduce positive, normalized operators $\sigmaB$ and $\zetaR$ and the maps $\tilde{\cT}$ and $\tilde{\cE}$ as defined in equations \eqref{intrtilt} and \eqref{intrtile} respectively and find
\begin{align*}
&\ED{{\left\| (\cT\otimes\cE)((\idx{U}{A} \otimes \idi{\ensuremath{\tilde{A}}}) \:\idx{\xi}{A\ensuremath{\tilde{A}}}\:(\idx{U}{A}^\dagger \otimes \idi{\ensuremath{\tilde{A}}}))\right\|}_1^2}\nonumber\\
&\leq\ED{{\left\| (\tilde{\cT}\otimes\tilde{\cE})((\idx{U}{A} \otimes \idi{\ensuremath{\tilde{A}}}) \:\idx{\xi}{A\ensuremath{\tilde{A}}}\:(\idx{U}{A}^\dagger \otimes \idi{\ensuremath{\tilde{A}}}))\right\|}_2^2}\\
&=\ED{\Trace{(\tilde{\cT}\otimes\tilde{\cE})((\idx{U}{A} \otimes \idi{\ensuremath{\tilde{A}}})  \:\idx{\xi}{A\ensuremath{\tilde{A}}}\: (\idx{U}{A}^\dagger \otimes \idi{\ensuremath{\tilde{A}}}))}}.
\end{align*}
Applying the swap trick and using the definitions of the adjoint mappings of $\tilde{\cT}$ and $\tilde{\cE}$ gives
\begin{align*}
&\ED{\Trace{(\tilde{\cT}\otimes\tilde{\cE})((\idx{U}{A} \otimes \idi{\ensuremath{\tilde{A}}})\:\idx{\xi}{A\ensuremath{\tilde{A}}}\:(\idx{U}{A}^\dagger \otimes \idi{\ensuremath{\tilde{A}}}))^2}}\nonumber\\
&=\ED{\Trace{\left((\idx{U}{A} \otimes \idi{\ensuremath{\tilde{A}}})^{\otimes2} \ (\idx{\xi}{A\ensuremath{\tilde{A}}})^{\otimes2} \ (\idx{U}{A}^\dagger \otimes \idi{\ensuremath{\tilde{A}}})^{\otimes 2}\right) (\tilde{\mathcal{T}}^{\dagger})^{\otimes2}[\idx{\mathcal{F}}{B}]\otimes (\tilde{\cE}^{\dagger})^{\otimes2}[\idx{\mathcal{F}}{R}]}}.
\end{align*} 
With the relations
\begin{align*}
\ED{\left((\idx{U}{A}^{\otimes2} \otimes \idi{\ensuremath{\tilde{A}}}^{\otimes2}) \ (\idx{\xi}{A\ensuremath{\tilde{A}}})^{\otimes2} \ ((\idx{U}{A}^\dagger)^{\otimes2} \otimes \idi{\ensuremath{\tilde{A}}}^{\otimes2})\right)}&=(\cG_W\otimes\opidi{\ensuremath{\tilde{A}\tilde{A'}}})(\idx{\xi}{A\ensuremath{\tilde{A}}}^{\otimes2}),\\
\EH{\left((\idx{U}{A}^{\otimes2} \otimes \idi{\ensuremath{\tilde{A}}}^{\otimes2}) \ (\idx{\xi}{A\ensuremath{\tilde{A}}})^{\otimes2} \ ((\idx{U}{A}^\dagger)^{\otimes2} \otimes \idi{\ensuremath{\tilde{A}}}^{\otimes2})\right)}&=(\cG_H\otimes\opidi{\ensuremath{\tilde{A}\tilde{A'}
}})(\idx{\xi}{A\ensuremath{\tilde{A}}}^{\otimes2}),
\end{align*}
we have:
\begin{align}
&\Trace{\ED{\left({(\idx{U}{A}^{\otimes2} \otimes \idi{\ensuremath{\tilde{A}}}^{\otimes2}) \ (\idx{\xi}{A\ensuremath{\tilde{A}}})^{\otimes2} \ ((\idx{U}{A}^\dagger)^{\otimes2} \otimes \idi{\ensuremath{\tilde{A}}}^{\otimes2})}\right)} (\tilde{\mathcal{T}}^{\dagger})^{\otimes2}[\idx{\mathcal{F}}{B}]\otimes (\tilde{\cE}^{\dagger})^{\otimes2}[\idx{\mathcal{F}}{R}]}\nonumber\\
&=\Trace{\left((\cG_W\otimes\opidi{\ensuremath{\tilde{A}\tilde{A'}}})(\idx{\xi}{A\ensuremath{\tilde{A}}}^{\otimes2})-\left(\cG_H\otimes\opidi{\ensuremath{\tilde{A}\tilde{A'}}}\right)(\idx{\xi}{A\ensuremath{\tilde{A}}}^{\otimes2})\right) (\tilde{\mathcal{T}}^{\dagger})^{\otimes2}[\idx{\mathcal{F}}{B}]\otimes (\tilde{\cE}^{\dagger})^{\otimes2}[\idx{\mathcal{F}}{R}]}\nonumber\\
&\quad+\Trace{\left(\cG_H\otimes\opidi{\ensuremath{\tilde{A}\tilde{A'}}}\right)(\idx{\xi}{A\ensuremath{\tilde{A}}}^{\otimes2})\  
(\tilde{\mathcal{T}}^{\dagger})^{\otimes2}[\idx{\mathcal{F}}{B}]\otimes 
(\tilde{\cE}^{\dagger})^{\otimes2}[\idx{\mathcal{F}}{R}]}\label{tterms}.
\end{align}
For now we fix our attention on the first term of Equation~\eqref{tterms}. Bounding this term gives
\begin{align}
&\Norm{\left((\cG_W\otimes\opidi{\ensuremath{\tilde{A}\tilde{A'}}})(\idx{\xi}{A\ensuremath{\tilde{A}}}^{\otimes2})-\left(\cG_H\otimes\opidi{\ensuremath{\tilde{A}\tilde{A'}}}\right)(\idx{\xi}{A\ensuremath{\tilde{A}}}^{\otimes2})\right) (\tilde{\mathcal{T}}^{\dagger})^{\otimes2}[\idx{\mathcal{F}}{B}]\otimes (\tilde{\cE}^{\dagger})^{\otimes2}[\idx{\mathcal{F}}{R}]}{1}\nonumber\\
&\leq\Norm{\left(\cG_W\otimes\opidi{\ensuremath{\tilde{A}\tilde{A'}}}-\cG_H\otimes\opidi{\ensuremath{\tilde{A}\tilde{A'}}}\right)(\idx{\xi}{A\ensuremath{\tilde{A}}}^{\otimes2})}{1}\Norm{ (\tilde{\mathcal{T}}^{\dagger})^{\otimes2}[\idx{\mathcal{F}}{B}]}{\infty}\Norm{(\tilde{\cE}^{\dagger})^{\otimes2}[\idx{\mathcal{F}}{R}]}{\infty}\nonumber\\
&\leq\Norm{\cG_W-\cG_H}{\diamond}\Norm{\idx{\xi}{A\ensuremath{\tilde{A}}}^{\otimes2}}{1}\Norm{(\tilde{\mathcal{T}}^{\dagger})^{\otimes2}[\idx{\mathcal{F}}{B}]}{\infty}\Norm{(\tilde{\cE}^{\dagger})^{\otimes2}[\idx{\mathcal{F}}{R}]}{\infty}\nonumber\\
&\leq4\delta\Norm{(\tilde{\mathcal{T}}^{\dagger})^{\otimes2}[\idx{\mathcal{F}}{B}]}{\infty}\Norm{(\tilde{\cE}^{\dagger})^{\otimes2}[\idx{\mathcal{F}}{R}]}{\infty}\label{zsmbstl},
\end{align}
where inequality \eqref{zsmbstl} uses the explicit form of $\idx{\xi}{A\ensuremath{\tilde{A}}}=\idx{\Phi}{A\ensuremath{\tilde{A}}}-\idx{\pi}{A}\otimes\idx{\pi}{\ensuremath{\tilde{A}}}$ and the definition of the $\delta$-approximate two-design. In the following steps we upper bound the term $\norm{(\tilde{\mathcal{T}}^{\dagger})^{\otimes2}[\idx{\mathcal{F}}{B}]}{\infty}$. Let $\idx{P}{AA'}^+$ be the projector corresponding to the biggest absolute eigenvalue of $(\tilde{\mathcal{T}}^{\dagger})^{\otimes2}[\idx{\mathcal{F}}{B}]$. The $\infty$-norm can then be rewritten as
\begin{align}
\Norm{(\tilde{\mathcal{T}}^{\dagger})^{\otimes2}[\idx{\mathcal{F}}{B}]}{\infty}=\Abs{\Trace{(\tilde{\mathcal{T}})^{\otimes2}[\idx{P}{AA'}^+]\idx{\mathcal{F}}{B}}}\label{infnrmbnd}.
\end{align}
To be able to apply the swap trick, we decompose $\idx{P}{AA'}^+$ into some basis: $\idx{P}{AA'}^+=\sum\limits_{i,j} c_{ij} \idx{\sigma}{A}^i\otimes \idx{\sigma}{A'}^j$. Without loss of generality we choose the coefficients $c_{ij}$ to be real. This gives:
\begin{align}
\Trace{(\tilde{\mathcal{T}})^{\otimes2}[\idx{P}{AA'}^+]\idx{\mathcal{F}}{B}}=\sum\limits_{i,j} c_{ij}\Trace{\tilde{\mathcal{T}}(\idx{\sigma}{A}^i)\tilde{\mathcal{T}}(\idx{\sigma}{A'}^j)}.
\end{align}
We rewrite $\tilde{\mathcal{T}}(\idx{\sigma}{A}^i)$ using the Choi-Jamio\l{}kowski representation of $\tilde{\cT}$
\begin{align}
&\sum\limits_{i,j} c_{ij}\Trace{(\tilde{\mathcal{T}}(\idx{\sigma}{A}^i)\tilde{\mathcal{T}}(\idx{\sigma}{A'}^j))}\nonumber\\
&=\idx{d}{A}^2\sum\limits_{i,j}c_{ij}\Trace{\ptrace{A}{\idx{\tilde{\omega}}{AB}\:(\idi{B}\otimes(\idx{\sigma}{A}^i)^\intercal)}\:\ptrace{A'}{\idx{\tilde{\omega}}{A'B}\:(\idi{B}\otimes(\idx{\sigma}{A'}^j)^\intercal)}}\nonumber\\
&=\idx{d}{A}^2\Trace{(\idi{A'}\otimes\idx{\tilde{\omega}}{AB})\:(\idi{A}\otimes\idx{\tilde{\omega}}{A'B})\: (\idi{B}\otimes(\idx{P}{AA'}^+)^\intercal)}\label{ttexpd}.
\end{align}
To obtain an upper bound for Equation~\eqref{ttexpd} we apply the following Lemma~\ref{corebound}.
\begin{lemma}\label{corebound}Let $\idx{\omega}{AB}\in\hermops{\idx{\cH}{AB}},\ \idx{\omega}{A'B}\in\hermops{\idx{\cH}{A'B}}$ and let $\idx{\rho}{AA'}\in\hermops{\idx{\cH}{AA'}}$, then
$$\Abs{\Trace{(\idi{A'}\otimes\idx{\omega}{AB})\:(\idi{A}\otimes\idx{\omega}{A'B})\: (\idi{B}\otimes\idx{\rho}{AA'})}}\leq\Trace{\idx{\omega}{AB}^2}\sqrt{\Trace{\idx{\rho}{AA'}^2}}$$
\end{lemma}%
The proof of this lemma will be given after concluding the proof of Theorem~\ref{dcpwthaldsg}. We use the fact that $(\idx{P}{AA'}^+)^\intercal$ is a rank one projector and get
\begin{align}
\Trace{(\idi{A'}\otimes\idx{\tilde{\omega}}{AB})\:(\idi{A}\otimes\idx{\tilde{\omega}}{A'B})\: (\idi{B}\otimes(\idx{P}{AA'}^+)^\intercal)}\leq\Trace{\idx{\tilde{\omega}}{A'B}^2}\label{lem4}.
\end{align}
This gives the bound
\begin{align}
\Norm{(\tilde{\mathcal{T}}^{\dagger})^{\otimes2}[\idx{\mathcal{F}}{B}]}{\infty}\leq\idx{d}{A}^2\Trace{\idx{\tilde{\omega}}{A'B}^2}\nonumber.
\end{align}
And identically we find that
\begin{align}
\Norm{(\tilde{\cE}^{\dagger})^{\otimes2}[\idx{\mathcal{F}}{R}]}{\infty}\leq\idx{d}{A}^2\Trace{\idx{\tilde{\rho}}{AR}^2}\nonumber.
\end{align}
Thus we obtain the desired bound for the first term of \eqref{tterms} using \eqref{zsmbstl}:
\begin{align}
&\Norm{\left((\cG_W\otimes\opidi{\ensuremath{\tilde{A}\tilde{A'}}})(\idx{\xi}{A\ensuremath{\tilde{A}}}^{\otimes2})-\left(\cG_H\otimes\opidi{\ensuremath{\tilde{A}\tilde{A'}}}\right)(\idx{\xi}{A\ensuremath{\tilde{A}}}^{\otimes2})\right) (\tilde{\mathcal{T}}^{\dagger})^{\otimes2}[\idx{\mathcal{F}}{B}]\otimes (\tilde{\cE}^{\dagger})^{\otimes2}[\idx{\mathcal{F}}{R}]}{1}\nonumber\\
&\leq4\delta\idx{d}{A}^4\:\frac{1}{\tr{[\idx{\omega}{A'B}]}}\Trace{\idx{\tilde{\omega}}{A'B}^2}\:\frac{1}{\tr{[\idx{\rho}{AR}]}}\Trace{\idx{\tilde{\rho}}{AR}^2}\label{sfhjlk}.
\end{align}
The only thing left is to evaluate the second term of \eqref{tterms}, but this term was already calculated as part of the proof of the decoupling theorem. It equals the term on the right hand side of \eqref{traceterm} and can be bounded using \eqref{uslm}:
\begin{align}
&\Trace{\left(\cG_H\otimes\opidi{\ensuremath{\tilde{A}\tilde{A'}}}\right)(\idx{\xi}{A\ensuremath{\tilde{A}}}^{\otimes2})\  (\tilde{\mathcal{T}}^{\dagger})^{\otimes2}[\idx{\mathcal{F}}{B}]\otimes (\tilde{\cE}^{\dagger})^{\otimes2}[\idx{\mathcal{F}}{R}]}\nonumber\\
&\leq\frac{1}{\tr{[\idx{\omega}{A'B}]}}\Trace{\idx{\tilde{\omega}}{A'B}^2}\:\frac{1}{\tr{[\idx{\rho}{AR}]}}\Trace{\idx{\tilde{\rho}}{AR}^2}\label{wntmin}
\end{align}
An application of Lemma~\ref{minentr} on \eqref{sfhjlk} and \eqref{wntmin} gives
\begin{align*}
&\ED{\norm{\mathcal{T}((\idx{U}{A} \otimes \idi{R}) \ \rhoAR \  (\idx{U}{A}^\dagger \otimes \idi{R})) - \idx{\omega}{B} \otimes \idx{\rho}{R}}{1}^2}\nonumber\\
&\leq\left(1+4\delta\idx{d}{A}^4\right)\:2^{-\chmin{A'}{B}{\omega} - \chmin{A}{R}{\rho}},
\end{align*}
which proves \eqref{mchop} and thus concludes the proof of the decoupling theorem with approximate two-designs.
\end{proof}
%
%
\begin{proof}[Proof of Lemma~\ref{corebound}]
	We introduce a basis $\{\idx{\sigma}{A}^i\}_{i}$ for $\hermops{\idx{\cH}{A}}$ and a basis $\{\idx{\sigma}{B}^i\}_{i}$ for $\hermops{\idx{\cH}{B}}$. Moreover we choose them to be orthonormal with respect to the Schmidt scalar product (i.e. $\tr(\sigma_A^i \sigma_A^j) = \delta_{ij}$ and likewise for the $B$ system). Hence, the product operators $\{\idx{\sigma}{A}^{i}\otimes\sigma^{j}_B\}_{i,\:j}$ also form an orthonormal basis for $\hermops{\idx{\cH}{AB}}$ with respect to the Schmidt scalar product: 
\begin{align*} \Trace{(\idx{\sigma}{A}^{i}\otimes\idx{\sigma}{B}^j)\:(\idx{\sigma}{A}^k\otimes\idx{\sigma}{B}^l)}=\Trace{\idx{\sigma}{A}^{i}\idx{\sigma}{A}^k}\cdot\Trace{\idx{\sigma}{B}^j\idx{\sigma}{B}^l}
=\delta_{ik}\delta_{jl}
\end{align*}
We write the operators $\idx{\omega}{AB}$, $\idx{\omega}{A'B}$ and $\idx{\rho}{AA'}$ in that basis:
\begin{align*}
\idx{\omega}{AB}\ &:=\ \sum\limits_{i,j}a_{ij}\idx{\sigma}{A}^{i}\otimes\idx{\sigma}{B}^j & a_{ij}\ &:=\ \Trace{(\idx{\sigma}{A}^{i}\otimes\idx{\sigma}{B}^j)\:\idx{\omega}{AB}},\\
\idx{\omega}{A'B}\ &:=\ \sum\limits_{i,j}a_{ij}\idx{\sigma}{A'}^i\otimes\idx{\sigma}{B}^j & a_{ij}\ &:=\ \Trace{(\idx{\sigma}{A'}^{i}\otimes\idx{\sigma}{B}^j)\:\idx{\omega}{A'B}},\\
\idx{\rho}{AA'}\ &:=\ \sum\limits_{i,j}c_{ij}\idx{\sigma}{A}^{i}\otimes\idx{\sigma}{A'}^j & c_{ij}\ &:=\ \Trace{(\idx{\sigma}{A}^{i}\otimes\idx{\sigma}{A'}^j)\:\idx{\rho}{AA'}}.
\end{align*}
Since all matrices in the above statements are hermitian, the coefficients $a_{ij}$ and $c_{ij}$ are real. Moreover the coefficients in the expansion of $\idx{\omega}{AB}$ and $\idx{\omega}{A'B}$ are the same, because the corresponding matrices are the same. Substituting the expansions into the left-hand side of the lemma gives:
\begin{align}
&\Trace{(\idi{A'}\otimes\idx{\omega}{AB})\:(\idi{A}\otimes\idx{\omega}{A'B})\:(\idi{B}\otimes\idx{\rho}{AA'})}\nonumber\\
&=\sum\limits_{i,j,k,l,m,n}{a_{ij}a_{kl}c_{mn}\Trace{(\idi{A'}\otimes\idx{\sigma}{A}^{i}\otimes\idx{\sigma}{B}^j)\:              (\idi{A}\otimes\idx{\sigma}{A'}^k\otimes\idx{\sigma}{B}^l)\: (\idi{B}\otimes\idx{\sigma}{A}^m\otimes\idx{\sigma}{A'}^n)}}\nonumber\\
&=\sum\limits_{i,j,k,l,m,n}{a_{ij}a_{kl}c_{mn}\Trace{\idx{\sigma}{A}^{i}\idx{\sigma}{A}^m}\Trace{\idx{\sigma}{A'}^k\idx{\sigma}{A'}^n}\Trace{\idx{\sigma}{B}^j\idx{\sigma}{B}^l}}\nonumber\\
&=\sum\limits_{i,j,k,l,m,n}{a_{ij}a_{kl}c_{mn}\delta_{im}\delta_{kn}\delta_{jl}}\nonumber\\
&=\sum\limits_{i,j,k}{a_{ij}a_{kj}c_{ik}}\label{intrmat}
\end{align}
We now introduce the matrices $A:=(a_{ij})$ and $C:=(c_{ij})$ and use Equation \eqref{intrmat} to find that
\begin{align}
\Abs{\Trace{(\idi{A'}\otimes\idx{\omega}{AB})\:(\idi{A}\otimes\idx{\omega}{A'B})\:(\idi{B}\otimes\idx{\rho}{AA'})}}&=\Abs{\Trace{A^\dagger C A}}\nonumber\\
&\leq \Norm{AA^\dagger}{1}\Norm{C}{\infty}\nonumber\\
&\leq \Trace{AA^\dagger}\Norm{C}{2}\label{frend}.
\end{align}
We calculate the Schatten 2-norm of $C$ using that $\Norm{C}{2}^2=\sum_{ij}{\abs{c_{ij}}^2}$ (\cite{Bhatia}) and the explicit formula for the $c_{ij}$:
\begin{align}
\Norm{C}{2}^2&=\sum_{ij}{\abs{c_{ij}}^2}\nonumber\\
&=\sum_{ij}{\Trace{(\idx{\sigma}{A}^{i}\otimes\idx{\sigma}{A'}^j)\:\idx{\rho}{AA'}}\Trace{(\idx{\sigma}{A}^{i}\otimes\idx{\sigma}{A'}^j)\:\idx{\rho}{AA'}}}\nonumber\\
&=\Trace{\left(\sum_{ij}{\Trace{\idx{\sigma}{A}^{i}\otimes\idx{\sigma}{A'}^j\idx{\rho}{AA'}}\idx{\sigma}{A}^{i}\otimes\idx{\sigma}{A'}^j}\right)\idx{\rho}{AA'}}\nonumber\\
&=\Trace{\idx{\rho}{AA'}^2}\label{prjd2}.
\end{align}
The trace term in \eqref{frend} can be calculated similarly. We use the explicit formula for the coefficients:
\begin{align}
\Trace{AA^\dagger}&=\sum_{ij}{a_{ij}a_{ij}}\nonumber\\
&=\sum_{ij}{\Trace{(\idx{\sigma}{A'}^{i}\otimes\idx{\sigma}{B}^j)\:\idx{\omega}{A'B}}\Trace{(\idx{\sigma}{A'}^{i}\otimes\idx{\sigma}{B}^j)\:\idx{\omega}{A'B}}}\nonumber\\
&=\Trace{\left(\sum_{ij}\Trace{\idx{\sigma}{A'}^{i}\otimes\idx{\sigma}{B}^j\idx{\omega}{A'B}}\idx{\sigma}{A'}^{i}\otimes\idx{\sigma}{B}^j\right)\idx{\omega}{A'B}}\nonumber\\
&=\Trace{\idx{\omega}{A'B}^2}.\label{conlem5.2}
\end{align}
Taking \eqref{prjd2} together with \eqref{conlem5.2} and substituting them into \eqref{frend} concludes the proof of Lemma~\ref{corebound}.%
\end{proof}
\subsection{A Smoothed Decoupling Formula for Approximate Two-Designs}
\label{decoup:smoo}\begin{sloppypar}
In order to achieve a tighter bound in the decoupling formula for approximate two-designs (Theorem~\ref{dcpwthaldsg}), we now introduce a modified upper bound stated in terms of \emph{smooth} conditional min-entropies (see Definition \ref{def:smooth-min-entropy}). We refer to \cite{DTHM} for a discussion of the optimality of decoupling in terms of these quantities. The smooth conditional min-entropy has the additional advantage that it reduces to the von Neumann entropy in the important special case where the state is a tensor product of many identical states, as shown by the Fully Quantum Asymptotic Equipartition Theorem (see Equation~\ref{QAEP}).\end{sloppypar}

\begin{theorem}\label{wialtwo} (Smoothed decoupling formula for $\delta$-approximate two-designs)
Let $\rhoAR\in\subnormstates{\idx{\cH}{AR}}$ be a subnormalized density operator and let $\cTAE$ be a linear map with Choi-Jamio\l{}kowski representation $\idx{\omega}{A'B}\in\subnormstates{\idx{\cH}{BA'}}$ and let $\varepsilon$ be such that $\min{\{\sqrt{\trace{\rho}},\sqrt{\trace{\omega}}\}}>\varepsilon\geq0$. Then
\begin{align}
&\ED{\left\|\mathcal{T}((\idx{U}{A} \otimes \idi{R}) \idx{\rho}{AR} (\idx{U}{A}^{\dagger} \otimes \idi{R}))-\idx{\omega}{B}\otimes\idx{\rho}{R}\right\|}_1\nonumber\\
&\qquad\qquad\qquad\qquad\leq\sqrt{1+4\delta\idx{d}{A}^4}\:2^{-\frac{1}{2}\:\chmineeps{\varepsilon}{A'}{B}{\omega}-\frac{1}{2}\:\chmineeps{\varepsilon}{A}{R}{\rho}}+8\idx{d}{A}\delta\:\varepsilon+12\varepsilon,\nonumber
\end{align}
where $\cD$ constitutes a $\delta$-approximate two-design.
\end{theorem}
\begin{proof}
Let $\idx{\hat{\omega}}{A'B}\in\subnormstates{\idx{\cH}{A'B}}$ be the state that saturates the bound in the definition of  $H_{\textnormal{min}}^\varepsilon$, i.e.\:$P(\idx{\omega}{A'B},\idx{\hat{\omega}}{A'B})\leq\varepsilon$ and $\chmin{A'}{B}{\hat{\omega}} = \chmineeps{\varepsilon}{A'}{B}{\omega}$. Analogously $\idx{\hat{\rho}}{AR}$ is defined to be an operator with $P(\idx{\hat{\rho}}{AR},\idx{\rho}{AR})\leq\varepsilon$ and $\chmin{A}{R}{\hat{\rho}} = \chmineeps{\varepsilon}{A}{R}{\rho}$.\\
Using inequality \eqref{lem:genfuchs}, we find that:
\begin{align}
\left\|\idx{\omega}{A'B}-\idx{\hat{\omega}}{A'B}\right\|_1 &\leq 2\varepsilon & \left\|\idx{\rho}{AR}-\idx{\hat{\rho}}{AR}\right\|_1 &\leq 2\varepsilon\label{omrobnd}.
\end{align}
We decompose $\hat{\omega}-\omega$ and $\hat{\rho}-\rho$ into positive operators with orthogonal support writing
\begin{align}
\hat{\omega}-\omega &= \Delta_+-\Delta_- & \hat{\rho}-\rho &= \Gamma_+-\Gamma_-\nonumber
\end{align}
and conclude from \eqref{omrobnd} that
\begin{align*}
\Norm{\Delta_+}{1}&\leq2\varepsilon & \Norm{\Delta_-}{1}&\leq2\varepsilon & \Norm{\Gamma_+}{1}&\leq2\varepsilon & \Norm{\Gamma_-}{1}&\leq2\varepsilon.
\end{align*}
Let $\hat{\cT}$, $\cD_+$ and $\cD_-$ be the unique Choi-Jamio\l{}kowski preimages of $\idx{\hat{\omega}}{A'B}$, $\Delta_+$ and $\Delta_-$ respectively. 
We apply Theorem~\ref{dcpwthaldsg} on $\hat{\rho}$ and $\hat{\omega}$ to find
\begin{align*}
&\sqrt{1+4\delta\idx{d}{A}^4}\:2^{-\frac{1}{2}\:\chmineeps{\varepsilon}{A'}{B}{\omega}-\frac{1}{2}\:\chmineeps{\varepsilon}{A}{R}{\rho}}\nonumber\\ &=\sqrt{1+4\delta\idx{d}{A}^4}\:2^{-\frac{1}{2}\:\chmineeps{}{A'}{B}{\hat{\omega}}-\frac{1}{2}\:\chmineeps{}{A}{R}{\hat{\rho}}}\\
&\geq\ED{{\left\| \mathcal{\hat{T}}((\idx{U}{A} \otimes \idi{R}) \ \idx{\hat{\rho}}{AR} \  (\idx{U}{A}^\dagger \otimes \idi{R})) - \idx{\hat{\omega}}{B} \otimes \idx{\hat{\rho}}{R} \right\|}_1}.
\end{align*}
For any unitary, we have with an application of the triangle inequality
\begin{align*}
&{\left\| \mathcal{\hat{T}}((\idx{U}{A} \otimes \idi{R}) \ \idx{\hat{\rho}}{AR} \  (\idx{U}{A}^\dagger \otimes \idi{R})) - \idx{\hat{\omega}}{B} \otimes \idx{\hat{\rho}}{R} \right\|}_1\\
&\geq{\left\| \mathcal{\hat{T}}((\idx{U}{A} \otimes \idi{R}) \ \idx{\hat{\rho}}{AR} \  (\idx{U}{A}^\dagger \otimes \idi{R})) - \idx{\omega}{B} \otimes \idx{\hat{\rho}}{R} \right\|}_1-2\varepsilon.
\end{align*}
In the same way $\idx{\hat{\rho}}{R}$ is eliminated from the product term and we get in total
\begin{align}
&{\left\| \mathcal{\hat{T}}((\idx{U}{A} \otimes \idi{R})\idx{\hat{\rho}}{AR}(\idx{U}{A}^\dagger \otimes \idi{R})) - \idx{\hat{\omega}}{B} \otimes \idx{\hat{\rho}}{R} \right\|}_1\nonumber\\
&\geq{\left\| \mathcal{\hat{T}}((\idx{U}{A} \otimes \idi{R}) \idx{\hat{\rho}}{AR} (\idx{U}{A}^\dagger \otimes \idi{R})) - \idx{\omega}{B} \otimes \idx{\rho}{R} \right\|}_1-4\varepsilon\nonumber\\
&\geq{\left\|\mathcal{T}((\idx{U}{A} \otimes \idi{R}) \idx{\rho}{AR} (\idx{U}{A}^{\dagger} \otimes \idi{R}))-\idx{\omega}{B}\otimes\idx{\rho}{R}\right\|}_1\nonumber\\
&\quad -{\left\|\mathcal{T}((\idx{U}{A} \otimes \idi{R}) \idx{\rho}{AR} (\idx{U}{A}^{\dagger} \otimes \idi{R}))-\mathcal{T}((\idx{U}{A} \otimes \idi{R}) \idx{\hat{\rho}}{AR} (\idx{U}{A}^{\dagger} \otimes \idi{R}))\right\|}_1\nonumber\\
&\quad- {\left\|\hat{\mathcal{T}}((\idx{U}{A} \otimes \idi{R}) \idx{\hat{\rho}}{AR} (\idx{U}{A}^{\dagger} \otimes \idi{R}))-\mathcal{T}((\idx{U}{A} \otimes \idi{R}) \idx{\hat{\rho}}{AR} (\idx{U}{A}^{\dagger} \otimes \idi{R}))\right\|}_1-4\varepsilon\label{finbrg}.
\end{align}
The first term of Equation \eqref{finbrg} corresponds to the unsmoothed decoupling formula.
For the remaining two terms
\begin{align}
\ED{{\left\|\mathcal{T}((\idx{U}{A} \otimes \idi{R}) \idx{\rho}{AR} (\idx{U}{A}^{\dagger} \otimes \idi{R}))-\mathcal{T}((\idx{U}{A} \otimes \idi{R}) \idx{\hat{\rho}}{AR} (\idx{U}{A}^{\dagger} \otimes \idi{R}))\right\|}_1}\label{frstsma}
\end{align}
and
\begin{align}
\ED{{\left\|\hat{\mathcal{T}}((\idx{U}{A} \otimes \idi{R}) \idx{\hat{\rho}}{AR} (\idx{U}{A}^{\dagger} \otimes \idi{R}))-\mathcal{T}((\idx{U}{A} \otimes \idi{R}) \idx{\hat{\rho}}{AR} (\idx{U}{A}^{\dagger} \otimes \idi{R}))\right\|}_1}\label{scndsma}
\end{align}
we need to find upper bounds. We treat them separately beginning with the first one.
To perform the calculation we write $\hat{\rho}-\rho=\Gamma_+-\Gamma_-$ and use the linearity of $\cT$. We get
\begin{align}
&\ED{{\left\|\mathcal{T}((\idx{U}{A} \otimes \idi{R}) \idx{\rho}{AR} (\idx{U}{A}^{\dagger} \otimes \idi{R}))-\mathcal{T}((\idx{U}{A} \otimes \idi{R}) \idx{\hat{\rho}}{AR} (\idx{U}{A}^{\dagger} \otimes \idi{R}))\right\|}_1}\nonumber\\
&\leq\sum_{a\in\{+,-\}}{\ED{{\left\|\mathcal{T}((\idx{U}{A} \otimes \idi{R}) \Gamma_a (\idx{U}{A}^{\dagger} \otimes \idi{R}))\right\|}_1}}\nonumber\\
&=\sum_{a\in\{+,-\}}\Trace{\mathcal{T}\Big(\left(\ED{}-\EH{}\right)\left((\idx{U}{A} \otimes \idi{R})\Gamma_a(\idx{U}{A}^{\dagger} \otimes \idi{R})\right)\Big)}\nonumber\\
&\quad+\sum_{a\in\{+,-\}}{\Trace{\cT\Big(\EH{\left((\idx{U}{A} \otimes \idi{R})\Gamma_a(\idx{U}{A}^{\dagger} \otimes \idi{R})\right)}\Big)}}\nonumber\\
&\leq\sum_{a\in\{+,-\}}\Norm{\left(\ED{}-\EH{}\right)\left((\idx{U}{A} \otimes \idi{R})\Gamma_a(\idx{U}{A}^{\dagger} \otimes \idi{R})\right)}{1}\Norm{\cT^\dagger(\idi{B})}{\infty}\nonumber\\
&\quad+\sum_{a\in\{+,-\}}\Trace{\mathcal{T}(\idx{\pi}{A})\otimes\ptr{A}{\Gamma_a}}\nonumber\\
&\leq\sum_{a\in\{+,-\}}{\delta\:\left\|\Gamma_a\right\|_1}{\left\|\cT^\dagger(\idi{B})\right\|}_{\infty} +\sum_{a\in\{+,-\}}\Trace{\idx{\omega}{A'B}}\Trace{\Gamma_a}\label{expl21dsg}\\
&\leq4\idx{d}{A}\delta\varepsilon+4\varepsilon\label{intprojdgr}.
\end{align}
Inequality \eqref{expl21dsg} used that an approximate two-design constitutes an approximate 1-design automatically. This can be seen straight from the definition by considering states that are given by the identity operator on one of the systems on which the unitaries act. The last inequality \eqref{intprojdgr} can be seen by choosing the eigenvalue of $\cT^\dagger(\idi{B})$ which is the biggest in absolute value and defining $\idx{P}{A}$ to be the projector corresponding to this eigenvalue. One then has ${\left\|\cT^\dagger(\idi{B})\right\|}_{\infty}\leq\idx{d}{A}$.\\
Bounding the term \eqref{scndsma} is done similarly. We decompose $\hat{\cT}-\cT\: =\: \cD_+-\cD_-$ in accordance with the decomposition $\hat{\omega}-\omega=\Delta_+-\Delta_-$. We then get
\begin{align}
&\ED{{\left\|\hat{\mathcal{T}}((\idx{U}{A} \otimes \idi{R}) \idx{\hat{\rho}}{AR} (\idx{U}{A}^{\dagger} \otimes \idi{R}))-\mathcal{T}((\idx{U}{A} \otimes \idi{R}) \idx{\hat{\rho}}{AR} (\idx{U}{A}^{\dagger} \otimes \idi{R}))\right\|}_1}\nonumber\\
&\leq\sum_{a\in\{+,-\}}{\Trace{\cD_a\Big(\ED{\:(\idx{U}{A} \otimes \idi{R})\:\idx{\hat{\rho}}{AR}\:(\idx{U}{A}^{\dagger} \otimes \idi{R})}\Big)}}\nonumber\\
&=\sum_{a\in\{+,-\}}{\Trace{\cD_a\Big(\left(\ED{}-\EH{}\right)\left((\idx{U}{A} \otimes \idi{R})\:\idx{\hat{\rho}}{AR}\:(\idx{U}{A}^{\dagger} \otimes \idi{R})\right)\Big)}}\nonumber\\
&+\sum_{a\in\{+,-\}}{\Trace{\cD_a\Big(\EH{\left((\idx{U}{A} \otimes \idi{R})\:\idx{\hat{\rho}}{AR}\:(\idx{U}{A}^{\dagger} \otimes \idi{R})\right)}\Big)}}\nonumber\\
&\leq\sum_{a\in\{+,-\}}{\Norm{\left(\ED{}-\EH{}\right)\left((\idx{U}{A} \otimes \idi{R})\:\idx{\hat{\rho}}{AR}\:(\idx{U}{A}^{\dagger} \otimes \idi{R})\right)}{1}}\Norm{\cD_a^\dagger(\idi{B})}{\infty}\nonumber\\
&+\sum_{a\in\{+,-\}}{\Trace{\cD_a(\idx{\pi}{A}\otimes\idx{\hat{\rho}}{R})}}\nonumber\\
&\leq\sum_{a\in\{+,-\}}\delta\:{\big\|\idx{\hat{\rho}}{AR}\big\|}_{1}\:{\left\|\cD_a^\dagger(\idi{B})\right\|}_{\infty}
+\sum_{a\in\{+,-\}}{\Trace{\Delta_a\otimes\idx{\hat{\rho}}{R}}}\nonumber\\
&\leq4\idx{d}{A}\delta\varepsilon +4\varepsilon.\label{scndtrm}
\end{align}
Combining the expressions \eqref{intprojdgr} and \eqref{scndtrm} and substituting them into \eqref{finbrg}, we obtain
\begin{align*}
&\ED{\left\| \mathcal{\hat{T}}((\idx{U}{A} \otimes \idi{R})\idx{\hat{\rho}}{AR}(\idx{U}{A}^\dagger \otimes \idi{R})) - \idx{\hat{\omega}}{B} \otimes \idx{\hat{\rho}}{R} \right\|}_1\nonumber\\
&\geq\ED{\left\|\mathcal{T}((\idx{U}{A} \otimes \idi{R}) \idx{\rho}{AR} (\idx{U}{A}^{\dagger} \otimes \idi{R}))-\idx{\omega}{B}\otimes\idx{\rho}{R}\right\|}_1-8\idx{d}{A}\delta\varepsilon-12\varepsilon.
\end{align*}
Finally this yields 
\begin{align*}
&\ED{\left\|\mathcal{T}((\idx{U}{A} \otimes \idi{R}) \idx{\rho}{AR} (\idx{U}{A}^{\dagger} \otimes \idi{R}))-\idx{\omega}{B}\otimes\idx{\rho}{R}\right\|}_1\nonumber\\
&\leq\sqrt{1+4\delta\idx{d}{A}^4}\:2^{-\frac{1}{2}\:\chmineeps{\varepsilon}{A'}{B}{\omega}-\frac{1}{2}\:\chmineeps{\varepsilon}{A}{R}{\rho}}+8\idx{d}{A}\delta\:\varepsilon+12\varepsilon,
\end{align*}
which proves the smoothed decoupling formula for $\delta$-approximate two-designs.
\end{proof}

\section{Decoupling in physical systems}
\label{decoup:anal}

In this section we explain how our result can be applied to study a typical evolution of a physical system. Consider, as before, a joint system $AR$ in an initial state $\rhoAR$ and assume that the $A$ system consists of a large number of interacting particles. {In a physical scenario $A$ might be correlated with a huge, diffuse subsystem of the universe such that $R$ might be much larger than $A$.} The most common type of interaction in nature is a local two-particle interaction. It can be modeled using a two-qubit unitary gate. More generally, one may describe the randomization process induced by the evolution of a many-particle system using a quantum circuit. Such approaches were considered earlier for instance in \cite{ETE} and \cite{RQC}. The circuit is constructed in the following way: at each step of the circuit, two qubits from $A$ and an element of a universal gate set for $\mathbb{U}(4)$ are chosen uniformly at random. The gate is applied to the qubits and the circuit proceeds to the next step. For a given circuit time $t$, we consider the set of all possible unitaries the circuit can produce together with the corresponding probabilities. If $t$ goes to infinity this yields the Haar distribution on the whole unitary group \cite{RQC}. Unfortunately, it turns out that the convergence rate of the random circuit towards the Haar distribution is exponentially slow in the number of qubits of the underlying system \cite{RQC,ETE,NielsenChuang}. Nevertheless, after a time $t$ that grows polynomially in the number of qubits and logarithmically in $\frac{1}{\delta}$, the above circuit will constitute a $\delta$-approximate two-design.

More precisely, the authors of \cite{RQC} (Theorems 2.9 and 2.10) and \cite{correct} derive the following pivotal theorem.
\begin{theorem}(Random quantum circuits are approximate two-designs, \cite{RQC,correct})\label{almst2dsglocqu} Let $\mu$ be the probability distribution corresponding to any universal gate set on $\mathbb{U}(4)$ and let $W$ be a random circuit on $n$ qubits obtained by drawing $t$ random unitaries according to $\mu$ and applying each of them to a random pair of qubits. Then there exists $C$ (and $C=C(\mu)$ only) 
such that for any $\delta>0$ and any $t\geq C(n^2+n\log(1/\delta))$, the set of unitaries produced by $W$ together with the corresponding probabilities forms a $\delta$-approximate unitary two-design.
\end{theorem}

Following the discussion in \cite{ETE}, we will assume that typical dynamics in nature are given by (short) circuits of the type of Theorem~\ref{almst2dsglocqu}. We conclude that in our model the possible evolutions of a many qubit system are given by elements of a unitary approximate two-design. Moreover, Theorem~\ref{almst2dsglocqu} states that in order to reach a $\delta$-approximate two-design a circuit time $t:= C(n^2+n\log{\frac{1}{\delta}})$ is sufficient, with $C$ being some constant that only depends on the concrete circuit used.

We can now apply our decoupling theorem for approximate two-designs to infer conditions under which typical processes in nature result in decoupling. 
In this example, we shall assume that the $R$ system is correlated with a subsystem of $A$ and we are interested in how this correlation behaves under a typical evolution. Hence, we decompose $A$ into two parts: $A_S$, which identifies the subsystem of interest, and $A_E$, which corresponds to an environmental system which is uncorrelated with $R$. Since we are interested in the state of $A_S$ we choose $\mathcal{T}$ to be the partial trace on the environment system: $\mathcal{T}(\rho) = \tr_{A_E}[\rho]$. Formally, this implies that $\chmin{A}{R}{\rho} \geq - \log d_{A_S}$ and $\chmin{A'}{E}{\omega} \geq \log d_{A_E} - \log d_{A_S}$ (see Lemma 20 in \cite{TCR09}). An application of Markov's inequality to the decoupling formula for approximate two-designs shows that, for any $\epsilon>0$, one has
\begin{align*}
&\Pr_W\Big\{{\norm{\ptrace{A_E}{(\idx{W}{A} \otimes \idi{R}) \:\rhoAR \:(\idx{W}{A}^\dagger \otimes \idi{R})} - \idx{\pi}{A_S} \otimes \idx{\rho}{R}}{1}}\geq \epsilon \Big\}\leq\\
&\qquad\qquad\frac{1}{\epsilon}\:\frac{d_{A_S}}{\sqrt{d_{A_E}}}\sqrt{1+4\delta d_A^4} .
\end{align*}
This implies that if the environment $A_E$ is chosen big enough, decoupling occurs except with small probability. Note, moreover, that the factor $\idx{d}{A}^4$ does not increase the time that is required until decoupling is reached in a significant way. To reach a $\bar{\delta}$-approximate two-design with $\bar{\delta}:= \frac{\delta}{\idx{d}{A}^4}$ it is sufficient to have run the circuit for a time
\begin{align*}
\bar{t} := C\left(n^2+n\log{\left(\frac{2^{4n}}{\delta}\right)}\right) =C\left(n^2+4n^2+n\log{\left(\frac{1}{\delta}\right)}\right)
\end{align*}

This means that once the circuit has reached a $\delta$-approximate two-design, it suffices to wait only approximately five times longer until it generates a $\bar{\delta}$-approximate two-design. This additional time certainly does not affect our conclusions.

We summarize our discussion with a corollary and give an outlook for possible applications of our results.
\begin{corollary}\label{finc}
Given a system $A$ which consists of two subsystems $A_S$ and $A_E$, assume that $A_S$ is correlated with a reference system $R$. Furthermore assume the $A$ system to consist of interacting particles, whose dynamics can be described with the above circuit model. Then if $A_E$ is chosen large enough a \textit{typical} process reaches decoupling after polynomial time except with small probability.
\end{corollary}

{In the context of black hole evaporation a result similar to Theorem~\ref{dcpwthaldsg} occurs in \cite{BLACK}, Inequality 5.1. However, the validity of this formula is restricted to the approximate 2-designs constructed in \cite{dankert09}, which share strong additional properties (\cite{dankert09}, Equation~(16)). In the model of \cite{BLACK} it seems reasonable to assume that the approximate 2-designs are generated via a random quantum circuit as in Corollary~\ref{finc}. Since in general such circuits will not produce the two designs of \cite{dankert09} our decoupling formula seems more appropriate for the application in \cite{BLACK} than Inequality 5.1. }

Finally, note that related results concerning the thermalization of subsystems have been derived in \cite{Lloyd:PHD,GemmerMahler,FoundStat} and a generalization of these results using the decoupling approach has recently been proposed in \cite{Hutter}.

\begin{acknowledgements}The authors would like to thank Mark Wilde for carefully reading the manuscript and pointing out some mistakes. This work was supported by the Swiss National Science Foundation (SNSF) through the National Centre of Competence in Research “Quantum Science and Technology” and through grant No.~200020-135048, and by the European Research Council through grant No.~258932.

\end{acknowledgements}
%
\bibliographystyle{abbrv}

\end{document}